\documentclass[twocolumn,prb,superscriptaddress,longbibliography]{revtex4-2}

\usepackage{amsmath,amsfonts,amssymb}
\usepackage{graphicx,color}
\usepackage{hyperref}
\usepackage{epstopdf}
\usepackage{float}
\usepackage{multirow}
\usepackage{hhline}
\usepackage{ulem}
\usepackage{mathtools}
\usepackage[caption=false]{subfig}
\usepackage{soul}

\usepackage{graphicx}
\usepackage{dcolumn}
\usepackage{bm}

\usepackage[utf8]{inputenc} 

\usepackage{multirow}
\usepackage[warn]{mathtext}
\usepackage{amssymb}

\usepackage{textcomp} 
\usepackage{indentfirst} 
\usepackage{amsmath} 
\usepackage{graphicx}
\DeclareGraphicsExtensions{.pdf,.png,.jpg}
\usepackage{pgfplots}
\pgfplotsset{compat=1.13}

\usepackage{hyperref}
\hypersetup{
    colorlinks=true,
    linkcolor=blue,
    filecolor=black,      
    urlcolor=blue,
    citecolor= violet
}

\usepackage{algpseudocode}

\usepackage{adjustbox}
\usepackage{tabularx}

\usepackage[nottoc]{tocbibind}
\usepackage{mathtools}

\usepackage{xcolor}
\definecolor{C0}{HTML}{4C72B0}
\definecolor{C1}{HTML}{DD8452}
\definecolor{C2}{HTML}{55A868}
\definecolor{C3}{HTML}{C44E52}



\newcolumntype{Y}{>{\centering\arraybackslash}X}

\makeatletter
\def\@fnsymbol#1{\ensuremath{\ifcase#1\or \dagger\or \ddagger\or
   \mathsection\or \mathparagraph\or \|\or **\or \dagger\dagger
   \or \ddagger\ddagger \else\@ctrerr\fi}}
\makeatother

\begin{document}

\title{High-fidelity transmon-coupler-activated CCZ gate on fluxonium qubits}

\author{Ilya A. Simakov$^*$}
\email{simakov.ia@phystech.edu}
\affiliation{Russian Quantum Center, 143025 Skolkovo, Moscow, Russia}
\affiliation{National University of Science and Technology ``MISIS'', 119049 Moscow, Russia}
\affiliation{Moscow Institute of Physics and Technology, 141701 Dolgoprudny, Russia}

\author{Grigoriy S. Mazhorin$^*$}
\affiliation{Russian Quantum Center, 143025 Skolkovo, Moscow, Russia}
\affiliation{National University of Science and Technology ``MISIS'', 119049 Moscow, Russia}
\affiliation{Moscow Institute of Physics and Technology, 141701 Dolgoprudny, Russia}

\author{Ilya N. Moskalenko}
\thanks{Present Address: Department of Applied Physics, Aalto University, Espoo, Finland}
\affiliation{National University of Science and Technology ``MISIS'', 119049 Moscow, Russia}

\author{Seidali S. Seidov}
\affiliation{National University of Science and Technology ``MISIS'', 119049 Moscow, Russia}

\author{Ilya S. Besedin}
\thanks{Present Address: Department of Physics, ETH Zurich, Zurich, Switzerland}
\affiliation{Russian Quantum Center, 143025 Skolkovo, Moscow, Russia}
\affiliation{National University of Science and Technology ``MISIS'', 119049 Moscow, Russia}

\date{\today}

\begin{abstract}

The Toffoli gate takes a special place in the quantum information theory. It opens up a path for efficient implementation of complex quantum algorithms. Despite tremendous progress of the quantum processors based on the superconducting qubits, realization of a high-fidelity three-qubit operation is still a challenging problem. Here, we propose a novel way to perform a high-fidelity CCZ gate on fluxoniums capacitively connected via a transmon qubit, activated by a microwave pulse on the coupler. The main advantages of the approach are relative quickness, simplicity of calibration and significant suppression of the unwanted longitudinal ZZ interaction. We provide numerical simulation of 95-ns long gate of higher than 99.99\% fidelity with realistic circuit parameters in the noiseless model and estimate an error of about 0.25\% under the conventional decoherence rates.

\end{abstract}


\maketitle

\def\thefootnote{*}\footnotetext{These authors contributed equally to this work}

\section{Introduction}

Three-qubit gates such as CCZ or Toffoli gate play an important role in quantum information theory \cite{Toffoli1980ReversibleC, Shi2003, aharonov2003simple}. They are relevant for numerous applications from the implementation of universal multi-controlled gates \cite{NielsenChuang} to the realization of the efficient quantum error correction protocols \cite{paetznick2013universal, yoder2016universal, reed2012realization}. The CCZ gate is a key ingredient for various algorithms of high practical importance such as variational quantum eigensolver \cite{peruzzo2014variational, kandala2017hardware}, Shor's algorithm \cite{10.5555/3179553.3179560}, approximate quantum optimization algorithm \cite{harrigan2021quantum}, quantum chemistry simulations \cite{lanyon2010towards}, which can be executed even on the noisy intermediate-scale quantum devices \cite{preskill2018quantum, RevModPhys.94.015004, figgatt2017complete}. For such applications a direct implementation of the three-qubit gate can significantly reduce the circuit depth in native hardware operations \cite{Lacroix_2020, Shi_2020} and improve the successful outcome of some algorithms \cite{Mariantoni_2011, Abrams_2020, PhysRevApplied.14.014072}.

Today, the quantum processors based on the superconducting transmon qubits evidence immense progress. The resent milestone experiments demonstrate tremendous coherence times increase \cite{Place2021, Wang2022}, quantum advantage \cite{arute2019quantum}, 
error suppression \cite{chen2021exponentia, google2023suppressing}, that altogether give good reasons to look ahead with optimism. On the superconducting platform there are several ways to implement a CCZ gate. The straightforward approach is to construct this three-qubit gate out of at least six two-qubit gates \cite{Shende2008OnTC}, which yields significantly lower three-qubit gate fidelity than the fidelity of the comprising single- and two-qubit gates. Another approach to improve the performance of CCZ gates is by using auxiliary levels of the qubits \cite{Fedorov2011, Nikolaeva_2022}. However, the intension to implement a direct three-qubit entangling gate is a preferable way which became a focus of a number of experimental research \cite{Fedorov2011, Warren_2023, reed2012realization}. The experimental work \cite{Kim_2022} shows the state-of-the-art 353-ns direct iToffoli operation of 98.26\% fidelity implemented by a simultaneous drive on three superconducting transmon qubits. The promising theoretical proposal in \cite{glaser2022controlledcontrolledphase} predict the three-qubit gate of less than 300~ns duration with fidelity above 99\%.

Superconducting fluxonium qubit \cite{Manucharyan_2009, Nguyen2019, Pop2014, PhysRevX.11.011010} becomes a promising alternative to the transmon as a base element of a quantum processor. A relatively low transition frequency, between 100 MHz and
1 GHz at the flux degeneracy point $0.5\Phi_0$, makes the fluxonium less sensitive to the dielectric loss. At the same time, a high (several GHz) anharmonicity presents an advantage for performing fast gates.
Energy relaxation times $T_1$ and coherence times $T_2$ exceeding 1 ms \cite{Pop2014, PhysRevLett.130.267001} made it possible to achieve single-qubit gate fidelities for fluxoniums above 99.99\%. Besides that, a variety of high-fidelity two-qubit gates relying on the direct capacitive coupling \cite{Ficheux2021, Bao_2021} or mediated via tunable coupler \cite{Moskalenko2021, Moskalenko2022, PRXQuantum_3_040336, ding2023highfidelity} were proposed and implemented on fluxonium qubits. All of the above gives confidence in the possibility of using fluxoniums for the development of large-scale quantum systems, and implementation of three-qubit entangling operations on the fluxonium-based devices could provide means for a fault-tolerant quantum computation based on the low frequency qubits.



\begin{figure*}
    \centering
    \includegraphics[width=\linewidth]{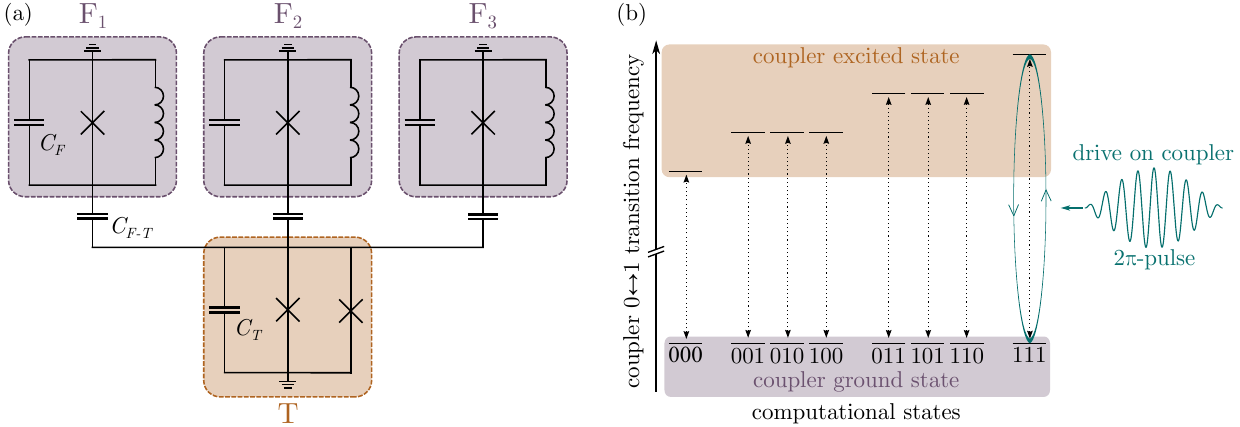}
    \caption{CCZ gate concept. (a) Circuit layout of the proposed device consisting of three computational fluxoniums F$_1$, F$_2$, F$_3$ (purple) capacitively coupled by a transmon T (orange). (b) The state-dependent spectrum of the coupler $0-1$ transition. After a single Rabi oscillation of the coupler population, caused by an external drive at the coupler main transition associated with the $|111\rangle$ computational state, the state $|111\rangle$ accumulates extra phase $\pi$. The rest of the computational states remains undisturbed, that altogether corresponds to the action of a CCZ operation.}
    \label{fig:circuit_concept}
\end{figure*}

Here, we propose a CCZ gate implementation using three fluxoniums capacitively coupled to an extra transmon qubit. Due to the strong interaction between the subsystems, the spectrum of the coupler is dependent on the data qubits' state. Thereby, driving the coupler transition associated with the chosen computational state, one can accumulate the phase exactly on this state. The concept is motivated by the experiments \cite{ding2023highfidelity, simakov2023coupler} where the microwave pulse on the coupling qubit is used for acquiring conditional phase of the two-qubit CZ operation. 
The gate has a number of notable advantages: it is simple for calibration, suppresses the unwanted longitudinal ZZ interaction, which along with decoherence is an important limiting factor for performance of superconducting qubits. Also, within the duration of the gate, the second and higher excited states of the qubits are only dispersively involved, while when the gate is not active, these levels are not involved at all. We present a numerical simulation of the proposed gate concept that shows the CCZ operation of 95~ns duration with fidelity above 99.99\% in the noiseless model and higher than 99.7\% with the conventional decoherence rates of the up-to-date superconducting qubits.


\section{Results}

\subsection{System concept}


\begin{table}[b]
    \centering
    \begin{tabularx}{\columnwidth}{@{}l *6{>{\centering\arraybackslash}X}@{}}
    \hline
    \hline
         & $E_L/h$, GHz & $E_C/h$, GHz & $C$, fF & $E_J/h$, GHz & $f_{01}$, GHz\\
         \hline
          \textbf{F1} & 1.2 & 1.53 & 12.88 & 6.35 & 0.576 \\
          \textbf{F2} & 1.2 & 1.53 & 12.88 & 6.25 & 0.598 \\
          \textbf{F3} & 1.2 & 1.53 & 12.88 & 6.15 & 0.621 \\
          \textbf{T} & - & 0.3 & 67.1 & 22.75 & 7.075 \\
    \hline
    \hline
    \end{tabularx}
    \begin{tabularx}{\columnwidth}{@{}l *3{>{\centering\arraybackslash}X}@{}}
         & F-T, MHz & F-F, MHz \\
         Coupling strength $g_{ij}/h$ & 600 & 150 \\
    \hline
    \hline
    \end{tabularx}
    \begin{tabularx}{\columnwidth}{@{}l *3{>{\centering\arraybackslash}X}@{}}
                  & F-T, fF \\
         Coupling capacitance $C$ & 3.22 \\
    \hline
    \hline
    \end{tabularx}
    \caption{Circuit parameters used throughout the work.}
    \label{tab:params}
\end{table}


\begin{figure}[b]
    \centering
    \includegraphics[width=\linewidth]{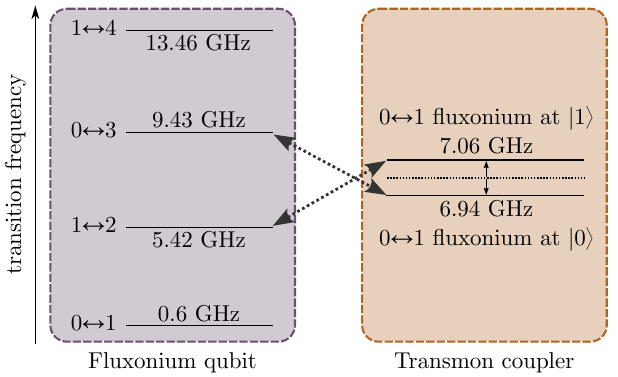}
    \caption{The schematic representation of the interaction between one of the three fluxonium qubits and transmon coupler. The left purple box represents the lowest allowed energy transitions from the $|0\rangle$ and $|1\rangle$ fluxonium states. In the right orange box the dashed line shows undisturbed transition frequency of the $0-1$ transmon transition. Due to the qubit interaction, illustrated by the arrows, the fluxonium transitions $1-2$ and $0-3$ push apart the transmon transition associated with fluxoniums states $|0\rangle$ and $|1\rangle$. As a result, the coupler transition frequency is different for the different computational states. By similar arguments, considering transmon coupler simultaneously interacting with three fluxonium qubits, we obtain a state-dependent spectrum of the coupler shown in Fig~\ref{fig:circuit_concept}b.
    }
    \label{fig:spectrum}
\end{figure}

\begin{figure}[t]
    \centering
    \includegraphics[width=\linewidth]{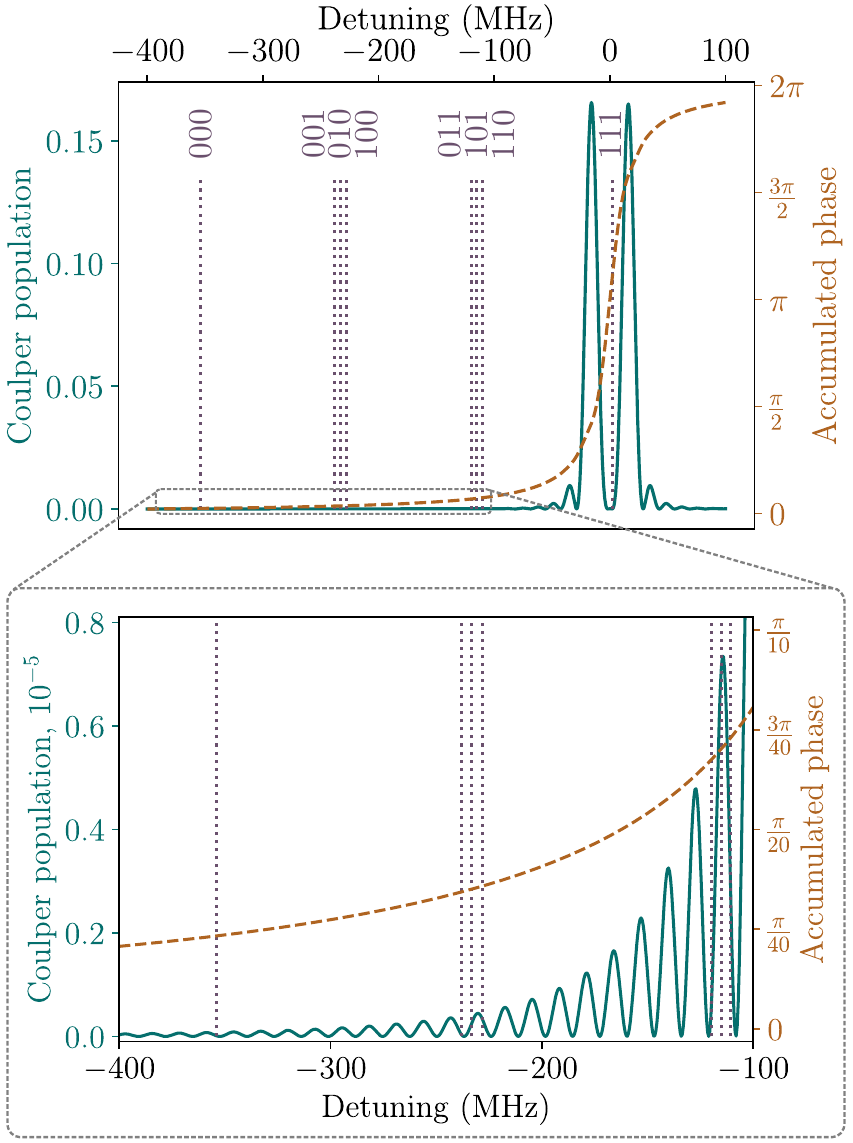}
    \caption{The detuning dependence of the coupler behaviour, approximated by a two-level model, after a $78$~ns-long  Gaussian $2\pi$-pulse.  The solid green line shows the system population (left axis) and the orange dashed curve presents the common phase acquired on the both states (right axis). The vertical dotted lines denote the location of the coupler $0-1$ transition associated with the eight computational states. The enlarged area shows the behavior of the population and the phase, accumulated on the qubit states, far from the resonance that is essential for the gate.}
    \label{fig:rabi_pops}
\end{figure}

\begin{table}[t]
    \centering    
    \begin{tabularx}{\columnwidth}{@{} *8{>{\centering\arraybackslash}X}@{}}
    \hline
    \hline
         $f_{000}$ & $f_{001}$ & $f_{010}$ & $f_{011}$ & $f_{100}$ & $f_{101}$ &$f_{110}$ & $f_{111}$ \\
         \hline
         6.9435 & 7.0596 & 7.0644 & 7.1777 & 7.0696 & 7.1825 & 7.1870 & 7.2994 \\
    \hline
    \hline
    \end{tabularx}
    \caption{The resulting transmon transition frequencies in GHz associated with different computational states.}
    \label{tab:freqs}
\end{table}

We consider a system of three low-energy fluxonium qubits capacitively coupled by a transmon, as schematically shown in Fig.~\ref{fig:circuit_concept}a. The computational qubits are biased at their degeneracy point $0.5\Phi_0$ and the coupler is set at zero flux. In the operating point, the coupler is deactivated and the residual ZZ interaction between computational qubits is small. The Hamiltonian of the full system can be expressed in the following form
\begin{equation}
    \mathcal{H}_\mathrm{full} = \sum_{i=\mathrm{F}_1, \mathrm{F}_2, \mathrm{F}_3, \mathrm{T}}  \mathcal{H}_i + \sum_{\substack{ i,j \in \left\{ \mathrm{F}_1, \mathrm{F}_2, \mathrm{F}_3, \mathrm{T} \right\} \\ i \neq j}} g_{ij}n_i n_j,
    \label{eq:H_3q}
\end{equation}
where $g_{ij}$ is the coupling strength, $\mathcal{H}_i$ is the Hamiltonian of the corresponding fluxonium \cite{Manucharyan_2009} and transmon qubits~\cite{PhysRevA.76.042319} 
\begin{equation}
    \mathcal{H}_\text{F} = 4E_C n^2 + E_J \left( 1 - \cos{\varphi} \right) + \frac{1}{2} E_L \left(\varphi - \frac{\pi}{2} \right)^2,
    \label{eq:H_fluxonium}
\end{equation}
\begin{equation}
    \mathcal{H}_\text{T} = 4E_C n^2 + E_{J} \left( 1 - \cos{\varphi} \right).
    \label{eq:H_transmon}
\end{equation} 
Here $n$ and $\varphi$ are the dimensionless charge and flux operators satisfying a commutation relationship $[n,\varphi]=i$, $E_C$, $E_L$ and $E_J$ are the charging, inductive and Josephson energies. The computational qubits are chosen to be 22 MHz different, as in practice they are not identical due to the limitations of the fabrication process. The circuit parameters used throughout the work are given in Table~\ref{tab:params}. The gate resilience to the design parameters is thoroughly discussed in section~\ref{discussion}.

It turns out that in such a configuration the coupler $0-1$ transition frequency depends on the state of the computational qubits.
As it is shown in Fig.~\ref{fig:spectrum}, the fluxonium transition frequencies $1-2$ and $0-3$ are designed to be the nearest ones to the fundamental transition frequency of the transmon.
Thereby, the transition $1-2$ shifts the transmon transition associated with the excited fluxonium state up, and the transition $0-3$ shifts the transmon transition associated with the ground fluxonium state down. Altogether, these interactions introduce a fluxonium state-dependent dispersive shift of the transmon frequency. 
We mention  that the other fluxonium transitions such as $1-4$ and $0-1$ also contribute to the splitting, but the influence of the $1-2$ and $0-3$ transitions is dominant.

The state-dependent spectrum of the coupler with the respect to all three fluxoniums is shown in Fig.~\ref{fig:circuit_concept}b. We calculate the transition frequencies (see Table~\ref{tab:freqs}) by numerical diagonalization of the considered system, taking into account the first ten energy levels of each subsystem. We denote computational states of the qubits with three-component ket-vectors $|\text{F}_1\text{F}_2\text{F}_3\rangle$ and the extended qubit-coupler Hilbert space states as $|\text{F}_1\text{F}_2\text{F}_3\text{T}\rangle$. The frequency difference between the $|1110\rangle - |1111\rangle$ and $|0000\rangle - |0001\rangle$ transitions has an order of magnitude of tens of MHz, which allows to resolve the transitions with excitation pulses as short as tens to hundreds of nanosecond.
The idea of the gate is similar to a microwave-activated two-qubit controlled-Z gate \cite{Ficheux2021}, where non-computational transitions $|10\rangle - |20\rangle$ and $|11\rangle - |21\rangle$ are used to acquire a geometric phase. Here, instead of employing high-energy levels of the data qubits, we utilize the first coupler excited state to accumulate the required phase.

\subsection{Single-pulse CCZ gate}

We propose to use the first excited state of the coupler as an auxiliary level to acquire a geometric phase on one of the computational states by driving the coupler at the frequency of the $|1110\rangle - |1111\rangle$ transition. The challenge here is to prevent the drive signal from affecting the $0-1$ transmon transitions associated with the rest seven computational states. To address this concern, we choose a Gaussian envelope for the drive of duration $\tau$:
\begin{equation}
    V(t) = A \left\{ \exp{\left( \cfrac{(t-\tau/2)^2}{2\sigma^2} \right) } -  \exp{\left( \cfrac{(\tau/2)^2}{2\sigma^2} \right) } \right\},
    \label{eq:pulse}
\end{equation}
where $A$ is an amplitude and the pulse length is truncated by $\sigma = 0.4 \tau$. We add the corresponding time-dependent drive term to the Hamiltonian~(\ref{eq:H_3q}) as
\begin{equation}
    \mathcal{H}_\mathrm{drive}(t) = V(t) \cdot \sin \left( 2\pi f t \right) \cdot n_\mathrm{T},
    \label{eq:drive}
\end{equation}
where $f$ is the drive frequency and $n_\mathrm{T}$ is the charge operator of the transmon qubit.

\begin{figure}[t]
    \centering
    \includegraphics[width=\linewidth]{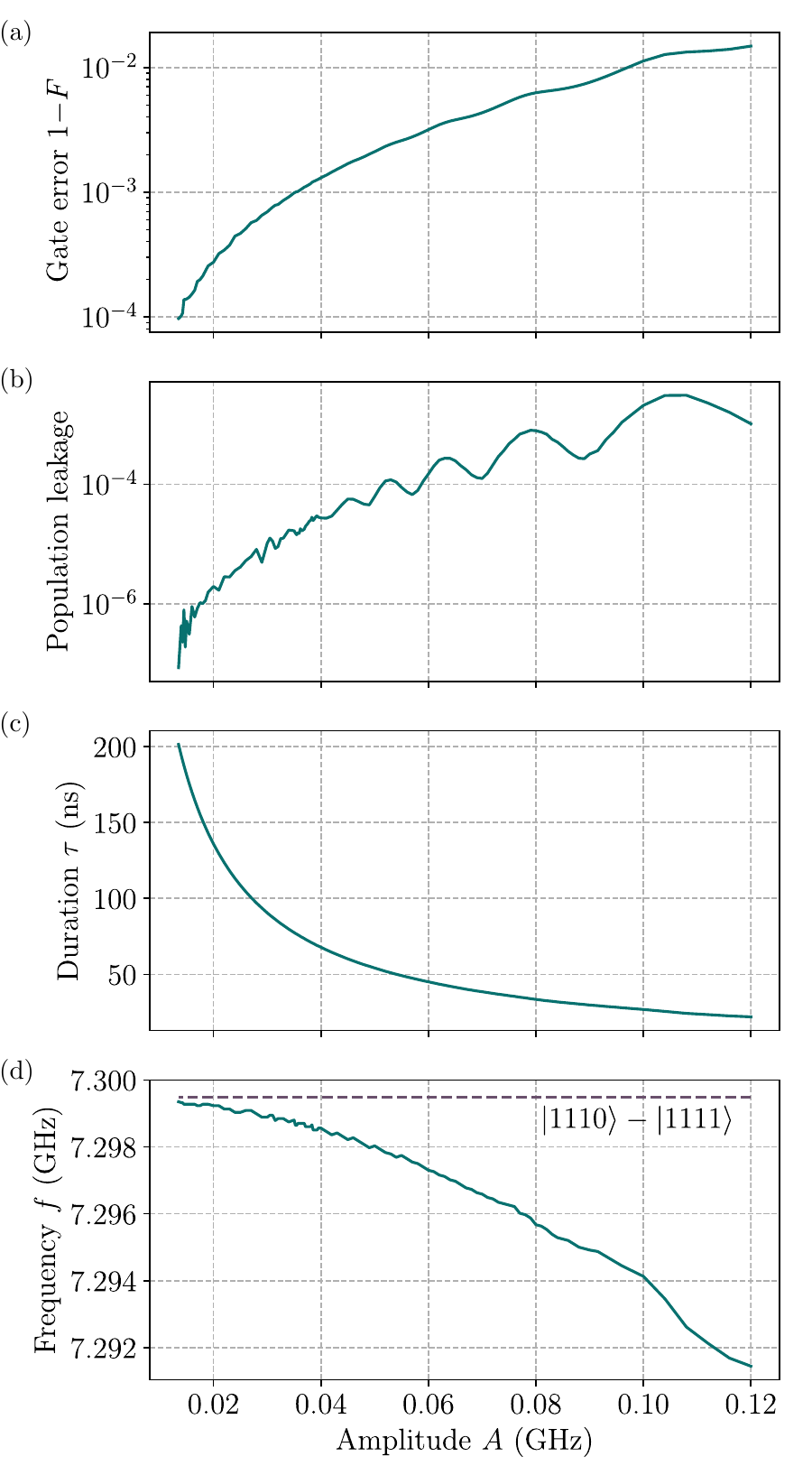}
    \caption{CCZ gate performance as a function of the Gaussian pulse amplitude. For each amplitude value we optimize gate fidelity as a function of the external pulse duration $\tau$ (c) and frequency $f$ (d), see (\ref{eq:pulse},~\ref{eq:drive}). The resulted optimal fidelity and corresponding to an average population leakage from the computational subspace associated only with the transmon ground state are shown in the plots (a) and (b). The bottom plot (d) demonstrates that with the decreasing pulse amplitude, the frequency becomes closer to the resonance transition $|1110\rangle - |1111\rangle$ frequency, indicated by a dashed line.
    }
    \label{fig:leakage}
\end{figure}

To grasp the gate concept, we simplify the proposed system and take into consideration only the first two levels of each of the four qubits in the same way as it is done in \cite{simakov2023coupler} and focus on the coupler transition. In this approximation, the system Hamiltonian can be separated into eight non-interacting two-level subsystems, each describing transmon $0-1$ transition associated with one of the eight fluxonium computational states as it shown in the spectrum in Fig.~\ref{fig:circuit_concept}b. The Hamiltonian of such two-level system can be expressed with the equation
\begin{equation}
    \frac{\widetilde{\mathcal{H}}}{h} = - \frac{\delta}{2}\sigma_z + \Omega(t) \sigma_x,
\end{equation}
where $\delta$ is a detuning between the two-level system frequency and the drive frequency, $\Omega(t)$ is the renormalized drive envelope~(\ref{eq:pulse}), $\sigma_x$ and $\sigma_z$ are the Pauli matrices. In Fig.~\ref{fig:rabi_pops} we 
display the behaviour of the two-level system after a Gaussian $2\pi$-pulse as a function of frequency detuning. The solid green line shows the system population and the orange dashed curve presents the common phase acquired on the both states.

Two conditions have to be satisfied simultaneously in order to perform a CCZ gate. First, at the end of the operation, the coupler should completely return to the initial state. Hence, the difference $\Delta=f_{111}-f_{110}$ between the closest transmon transitions has to be sufficiently large and at the same time the excitation pulse has to be long enough so that the residual coupler population after the gate can be considered negligible. 
Second, the $|111\rangle$ computational state should acquire a relative phase of $\pi$ compared to all other states. The orange dashed line in Fig.~\ref{fig:rabi_pops} shows the geometric phase accumulated after the drive pulse for varying detuning of the excitation pulse and the computational state-dependent transmon frequency.
All computational states acquire phases according their detuning. 
Three phases corresponding to single-qubit rotations can be compensated by a frame update \cite{McKay20177}, and the global phase can similarly be neglected, leaving 4 non-trivial phase accumulations corresponding to three pairs of two-qubit phases and a single three-qubit phase.
The single-qubit frame rotations also affect the target phase of the $|111\rangle$ state. However, due to the rapid growth of the accumulated phase nearby the resonance, slightly changing the signal detuning, one can compensate the effect of these additional phases.
We illustrate it in Fig.~\ref{fig:rabi_pops}: the drive tone is slightly detuned from the transition frequency associated with the $|111\rangle$.
The three two-qubit phases related to the double-excited computational states $|011\rangle$, $|101\rangle$, and $|110\rangle$ decrease when the pulse length is increased. At the same time, the optimal drive frequency approaches the  $|1110\rangle-|1111\rangle$ transition frequency. 

\begin{table*}
    \centering
    \begin{tabularx}{\textwidth}{@{} YYYYYYYYY @{}}
        \hline
        \hline
        CCZ & Gate duration, ns & Noiseless fidelity $F$, \% & Data qubits relaxation $T_1 = 300 \; \mu$s $\Delta F$, \% & Data qubits dephasing $T_\varphi = 100 \; \mu$s $\Delta F$ (\%) & Coupler relaxation $T_1 = 50 \; \mu$s $\Delta F$, \% & Coupler dephasing $T_\varphi = 50 \; \mu$s $\Delta F$, \% & Relaxation and dephasing in all qubits $\Delta F$, \%\\ 
        \hline
        Single-pulse & 78 & 99.90 & 0.035 & 0.098 & 0.005 & 0.058 & 0.198 \\
        Single-pulse & 195 & 99.99 & 0.086 & 0.250 & 0.013 & 0.151 & 0.500 \\
        Two-pulse & 95 & 99.99 & 0.043 & 0.121 & 0.005 & 0.076 & 0.245 \\
        \hline
        \hline
    \end{tabularx}
    \caption{Decoherence effect on the directly implemented CCZ operation and the two-pulse CCZ constructed by the quantum circuit in Fig.~\ref{fig:CCZ_calibration}. The 4-7th columns correspond to the gate error in case of only one relaxation channel acting simultaneously on all the computational qubits or only on the coupler. The last column shows the total gate infidelity conditioned by the decoherence in all elements of the considered system. Also, we mention that the two-pulse CCZ gate duration does not include time required for the single-qubit X rotations. The error rates are chosen to be close to the up-to-date conventional devices.}
    \label{tab:results}
\end{table*}

We simulate the time dynamics of the full system described by the Hamiltonian~(\ref{eq:H_3q}) under the proposed drive. We numerically integrate the Schrödinger equation for the eight computational states and obtain matrix representation of the three-qubit gate, which is non-unitary as we take into account leakage out the computational subspace. The details of the calculation are given in Appendix~\ref{app:methods}.
We fix an amplitude of the drive signal, and use the standard grid search technique to find optimal frequency and duration to get the largest gate fidelity. The steps of the duration and frequency sweep are $0.1$ ns and $0.1$ MHz, correspondingly.
Also we calculate the mean population over eight computational states related only to the ground state of the coupler. The optimal drive frequency and pulse length are shown in~Fig.~\ref{fig:leakage}, together with the corresponding gate fidelity and leakage. In addition to an exponential suppression of leakage with decreasing amplitude we observe an oscillatory dependence of both leakage and fidelity on amplitude. Leakage minima occur when the population of the coupler after the gate reaches a minimum not only for the triply-excited state $|111\rangle$, but also for doubly-excited states $|011\rangle$, $|101\rangle$, $|110\rangle$, see Fig.~\ref{fig:rabi_pops}.
As the drive amplitude decreases, the coupler population peak period also decreases.
Hence, we observe an oscillatory amplitude dependence of the population leakage in Fig.~\ref{fig:leakage}b.

\subsection{ZZ suppression protocol}

\begin{figure}[b]
    \centering
    \includegraphics[width=\linewidth]{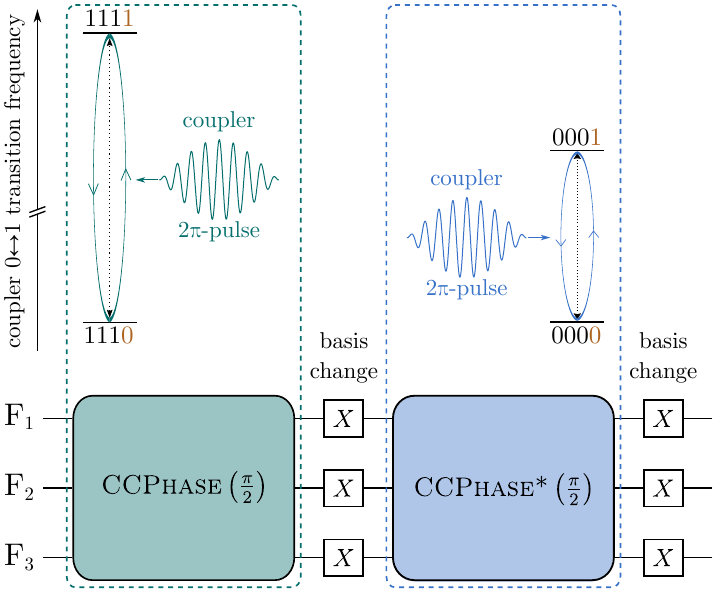}
    \caption{Quantum circuit of the two-pulse CCZ operation. CCPhase and CCPhase* gates are supposed to be implemented via driving the $|1110\rangle-|1111\rangle$ and $|0000\rangle-|0001\rangle$ transitions. The microwave $2\pi$-pulses on the transmon coupler are shown above the corresponding gates. The change of the computational basis allows one to cancel the unwanted longitudinal ZZ interaction with different phase accumulation signs during the CCPhase-like gates.}
    \label{fig:CCZ_calibration}
\end{figure}

According to the plots shown in Fig.~\ref{fig:leakage}, the population leakage
is not the dominant error source in the noiseless simulation. The major problem is an undesired longitudinal ZZ interaction, expressed by accumulation of phases of the double-excited computational states.

To deal with these unwanted processes, we notice that due to the rapid growth of the accumulated phase near the resonance, by varying the detuning of the drive signal one can get not only $\pi$, but choose an effective phase in a quite wide range about $\pi$ and consequently obtain a controlled-controlled-phase gate. Since there are two isolated transition $|1110\rangle - |1111\rangle$ and $|0000\rangle - |0001\rangle$ in the coupler spectrum (see. Fig.~\ref{fig:circuit_concept}b), the CCPhase operation can be implemented via the both transitions. 

Thereby, we propose to execute a CCPhase gate with phase $\pi/2$ via the transition $|1110\rangle - |1111\rangle$, then change the initial computational basis by three X rotations, implement the three-qubit $\text{CCPhase*}(\varphi)=I-|000\rangle \langle 000| (1-e^{i\varphi})$ gate with phase $\pi/2$ via the transition $|0000\rangle - |0001\rangle$, and finally return to the computational basis. The corresponding manipulation scheme is provided in Fig.~\ref{fig:CCZ_calibration}a. During the CCPhase and CCPhase* gates implementation the phases of the double-excited states during acquire different signs and thus suppress each other.

We simulate the time dynamics of the proposed circuit by the Schrödinger equation, assuming ideal single-qubit X rotations. The found CCPhase-like gates frequencies are $6.9284$ and $7.2835$ GHz, durations are $40.6$ and $54.3$ ns for the $\text{CCPhase*}(\pi/2)$ and $\text{CCPhase}(\pi/2)$ operations, correspondingly. The parasitic phases $\varphi_{011}$, $\varphi_{101}$, $\varphi_{110}$ are $0.031\pi$, $0.028\pi$, $0.027\pi$ and after implementation of both these gates in a row the remaining unwanted phases associated with the double excited computational states are less than $10^{-3}\cdot \pi$. As a result, we obtain a two-pulse CCZ gate of 95~ns duration with higher than 99.99\% fidelity.


\begin{figure*}
    \centering
    \includegraphics[width=\linewidth]{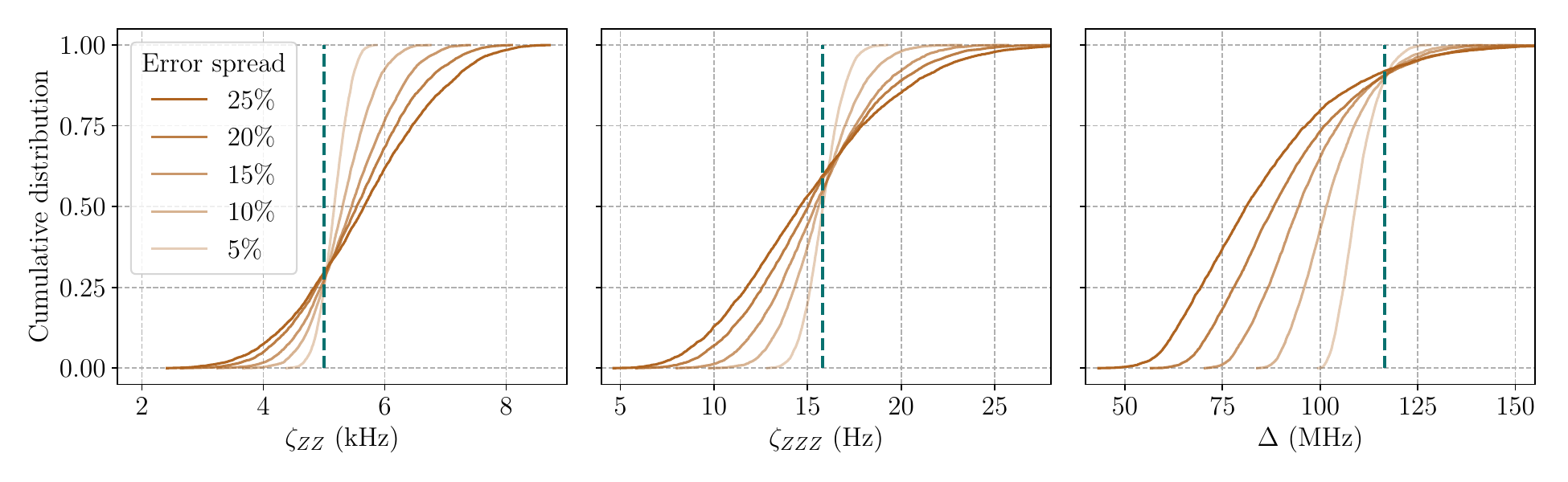}
    \caption{The cumulative distribution of the parasitic longitudinal interactions $\zeta_{ZZ}$, $\zeta_{ZZZ}$ and the characteristic value $\Delta$ of the state-dependence coupler spectrum calculated for different error spreads $\varepsilon$. The vertical line correspond to the designed parameters: $\zeta_{ZZ} = 5$~kHz, $\zeta_{ZZZ} = 15.8$~Hz, and $\Delta = 116$~MHz.}
    \label{fig:fun_prob}
\end{figure*}

Finally, we estimate the system performance under the decoherence effect for both single-pulse and two-pulse CCZ realization using the Lindblad master equation (see Appendix~\ref{app:methods} for calculation details). We separately consider the energy relaxation and dephasing on the computational fluxoniums and on the transmon coupler for the conservative relaxation rates \cite{PhysRevX.11.011010, Krinner_2022}. The results are collected in Table~\ref{tab:results}. We note that the dominant impact to the CCZ gate fidelity is exerted by the dephasing of the computational qubits and the coupler. Relaxation of the coupler degree of freedom, however, has a relatively low effect on the gate fidelity, because in all states but the $|111\rangle$ state the drive is off-resonant with the coupler. Also, we observe, that accounting both coherent and decoherent errors, the two-pulse gate demonstrates better fidelity compared to the single-pulse realization.

\section{Discussion}
\label{discussion}

It is crucial to discuss the attainability of the optimal circuit parameters of for the proposed CCZ gate implementation. As previously mentioned, the key feature of the system is the state-dependent coupler transition frequency. A large frequency difference~$\Delta$ between the transition from $|1110\rangle$ to $|1111\rangle$ and its nearest neighbouring transition of the coupler spectrum leads to a fast and consequently effective gate. Nevertheless, large coupling may cause the substantial parasitic longitudinal interaction. To effectively preserve the computation state while idling, one needs low two- and three-qubit unwanted interaction between computational qubits. Thus, the designed couplings should yield $\Delta$ that is large enough for a fast high-fidelity operation and the parasitic longitudinal $\zeta_{ZZ}$ and $\zeta_{ZZZ}$ interactions are sufficiently low compared to the decoherence rate.

Yet another significant benefit of the proposed system is its robustness with respect to fabrication imperfections. We qualitatively estimate the influence of critical current deviations to the system. The parameters of interest are the parasitic longitudinal $\zeta_{ZZ}$ and $\zeta_{ZZZ}$ interactions and the characteristic value $\Delta$ of the state-dependent coupler spectrum. We assume that the critical current is independently uniformly distributed within a range of $\pm \varepsilon$ of the designed value for each qubit. Using the Monte Carlo method we generate 10000 random samples of the critical current values, that randomly influence on the Josephson and inductive energies. This way, we collect the cumulative distributions of the parameters $\zeta_{ZZ}$, $\zeta_{ZZZ}$ and $\Delta$ for the different error spreads $\varepsilon$ and present them in Fig.~\ref{fig:fun_prob}. The dashed lines correspond to the designed values. The results indicate that the longitudinal two- and three-qubit interactions are small enough and lower $\Delta$ values can be offset by longer gate durations. Hence, we conclude that system is robust with respect to small variations of Josephson junction critical currents.

\section{Conclusion}

To summarize, in this work we propose a high-fidelity three-qubit CCZ gate implementation on fluxonium qubits capacitatively connected by a transmon coupler. The numerical simulation predicts the fidelity of the operation above 99.99\% with the duration less than 100~ns.  
The gate is activated by an external microwave drive close to the main transition frequency of the coupler 
relying on the same instrumentation as a regular single-qubit
gate on the transmon. Our approach has a number of significant advantages: when the coupler is de-excited, the spurious interaction between the computational states of the qubits is low. The leakage into non-computation states of the qubits can be strongly suppressed by increasing the gate length.
From an experimental point of view, the simplicity of the proposed gate implementation has evident benefits. The control pulse is applied only to the coupler, has no baseband component, while the microwave component has a conventional Gaussian envelope and only three essential parameters for calibration: duration, amplitude and frequency. 
The key feature of the gate concept is an ability to suppress the undesired longitudinal ZZ interaction by two sequential three-qubit controlled-phase gates via the coupler main transition associated with the ground  $|000\rangle$ and fully excited $|111\rangle$ states of the computational qubits. 

Moreover, the current method can be extended to the multi-controlled Z gate, allowing efficient entanglement of many qubits. Certainly, in this case, the number of coupled qubits will be limited by the coupling strength which cannot remain the same as the number of qubits increases due to the topology of the coupling structure. It is also worth noting additional challenges that might arise with more qubits. For example, the frequency crowding will set more limitations for the coupler drive pulses and moreover, the necessity to deal with parasitic phase accumulation will arise due to the additional three or more qubit ZZ interactions.
Wherein, the unwanted multi-qubit longitudinal interaction can be suppressed by the same procedure depicted in Fig.~\ref{fig:CCZ_calibration}.
Nevertheless, we believe the proposed gate principles open up a path toward the advanced architecture of superconducting qubit-based devices.

\section*{Acknowledgments}
The authors acknowledge Alexey Ustinov and Natalia Maleeva for helpful discussions and comments on the manuscript.
The work was supported by Rosatom in the framework of the Russian Roadmap for Quantum computing (Contract No. 868-1.3-15/15-2021 dated October 5, 2021).
I.A.S., G.S.M. and S.S.S. also thank the support of the Russian Science Foundation (Grant No. 23-72-01067).

\section*{Author contributions}
I.A.S., G.S.M., I.N.M., and I.S.B. conceived the ideas and developed the theory. I.A.S. performed dynamic simulations of the system. G.S.M. provided spectrum analysis.
All authors reviewed, discussed the results and contributed towards writing the manuscript.
I.A.S. and G.S.M. contributed equally to this work.


\appendix

\section{Simulation details}
\label{app:methods}

The Hamiltonian~(\ref{eq:H_fluxonium}) of each individual fluxonium is written in the phase basis, then the spectrum, charge and flux matrix elements are calculated by the Runge-Kutta method. The full circuit Hamiltonian~(\ref{eq:H_3q}) is constructed taking into account the first ten energy levels of each subsystem. Then the system is diagonalized and cut off to the first 128 energy levels, that compose the effective Hamiltonian used in the further research. The chosen number of energy levels is enough to account possible leakage to the second excited state of each subsystem. We use 128 energy levels to simulate the single-pulse three-qubit gate and only 32 levels for the two-pulse protocol and Lindblad evolution calculations.

We simulate the time evolution of the system in the absence of noise by numerically solving the Schrödinger equation with the computational stationary states $|\psi_i\rangle$ as the initial conditions. The resulted states $|\psi'_i\rangle$ are projected on the basis states and the gate operation matrix constructs as follows: $U_{ij} = \langle \psi_i | \psi'_j \rangle$. Then we choose an overall phase factor equal to zero so, that the element $U_{00}$ is real. Also, we apply single-qubit rotations about the $Z$-axis to set to zero phases on the $|001\rangle$, $|010\rangle$, and $|100\rangle$ states. Note that the method takes into account the leakage to the non-computational states and in general the matrix $U$ is not unitary. Thereafter, we transform the operator to the Pauli transfer matrix (PTM) $R$ and calculate the process fidelity by formula \cite{Nielsen_2002}
\begin{equation}
    F = \frac{\mathrm{Tr}(R^\dagger_\text{ideal}R) + 2^n}{2^n(2^n+1)},
    \label{eq:fidelity}
\end{equation}
where $n=3$ is the number of data qubits and $R_\text{ideal}$ is the PTM of the ideal gate.

To compute the time dynamics of the system with decoherence processes we use the Lindblad master equation:
\begin{equation}
    \dot\rho = -i[H, \rho] + \cfrac{1}{2} \sum_i \left( 2L_i \rho L_i^\dagger - \rho L_i^\dagger L_i - L_i^\dagger L_i \rho \right),
\end{equation}
where $L_i$ are the collapse operators. The corresponding operators to relaxation and dephasing are defined as 
\begin{equation}
    L_1 =\frac{1}{2\sqrt{T_1}} (\sigma_x + i \sigma_y), \; L_\varphi = \frac{1}{\sqrt{2T_\varphi}} \sigma_z,
\end{equation}
acting on a qubit. Here $\sigma_i$ are the Pauli matrices, padded with zeros to match the dimension of the corresponding subsystem. These operators are also transformed to the diagonal basis of the system. As we do not observe excitations of the high-energy levels of the system, we investigate relaxation processed related to the first excited state of the qubits and the coupler. Solving this equation for different initial states, one can obtain a superoperator of the corresponding noisy process \cite{wood2015tensor, nambu2005matrix}, convert it to the PTM representation and calculate the process fidelity with~(\ref{eq:fidelity}).


\normalem{}
\bibliography{main}

\end{document}